# Business intelligence systems and user's parameters: an application to a documents' database

**Babajide AFOLABI and Odile THIERY**
Laboratoire Lorrain de Recherche en Informatique et ses Application (LORIA)
Campus Scientifique BP 239
54506 Vandoeuvre-lès-Nancy, France.
babajide.afolabi@loria.fr, odile.thiery@loria.fr

**Abstract**

This article presents earlier results of our research works in the area of modeling Business Intelligence Systems. The basic idea of this research area is presented first. We then show the necessity of including certain users' parameters in Information systems that are used in Business Intelligence systems in order to integrate a better response from such systems. We identified two main types of attributes that can be missing from a base and we showed why they needed to be included. A user model that is based on a cognitive user evolution is presented. This model when used together with a good definition of the information needs of the user (decision maker) will accelerate his decision making process.

**Keywords:** Business intelligence, data warehouse, user modelling, data marts, informational needs, decisional needs.

## 1. Definitions: BIS and SIS

According to [Revelli, 1998], Business Intelligence (BI) "*is the process of collection, treatment and diffusion of information that has as an objective, the reduction of uncertainty in the making of all strategic decisions*".

In its simplest form, a strategic information system (SIS) can be considered as an information system (IS) consisting of "*strategic information and permits the automation of the organisation to better satisfy the objectives of the management*". For instance, an IS that aids in the management of stocks, we denote this as SI-S. A SIS can also be seen as "*an IS that is dedicated to strategic decision making and contains only strategic type of information*". For example, an IS that permits the decision maker to observe sales by country for a number of years or that permits an information watcher to point up the choices made during the analysis of the result obtained from an information search on the web. This is denoted as S-IS. [Tardieu and Guthmann, 1991; David and Thiery, 2003].

The following figure (figure 1) show that the organisation's IS is the first to be constructed. This IS is diverse and varies. It contains strategic information, for instance, indications on the organisation's turnover. From this then is extracted information necessary to the decision making process. Their structure (metadata) should also be extracted. This then constitutes the relational data warehouse, thus called because it is managed by a relational database management system (DBMS). Also, from this data warehouse is extracted multidimensional databases, which allows a view of the organisation from different angles or dimensions (for example, time axis, quantity of products sold or turnover). These multidimensional databases constitute the second type of SIS in the preceding paragraph. Their only contents are data that are necessary to the decision making process [Thiery et al., 2004].

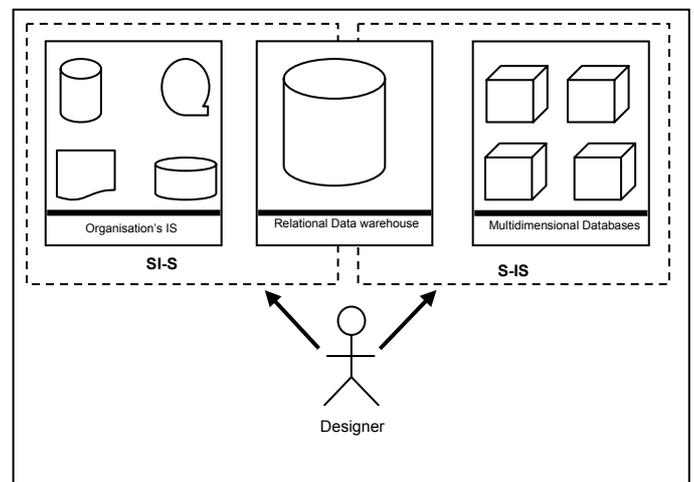

**Figure 1:** SI-S and S-IS

The decisions taken, using an IS, are based on the information found in it (the IS) and is also based on the user that has as an objective the appropriation of such system for a decision making process. For us, a Business Intelligence System BIS is a system that combines strategic information systems and user modelling domains. The final goal of a BIS is to help the user or the decision maker in his decision making process.

Figure 2 shows the architecture of a BIS in a process as proposed by the research team SITE[1]; one can easily identify the following four stages:

- **Selection:** selection permits the constitution of the IS of the organisation that can be (i) the production database (that allows current usage of the organisation), (ii) all the information support for an information retrieval system (in documentation for an example) or (iii) a SIS based on a data warehouse. This information system is constituted from heterogeneous data and from heterogeneous sources with the aid of a filter.
- **Mapping:** mapping permits all users an access to the data in the IS. Two methods of access are opened to the user: access by exploration and access by request. The exploration is based on a system of hypertexts. The requests are expressed with the aid of Boolean operators. The result of the mapping is a set of information.
- **Analysis:** in order to add value to the information found, techniques of analysis are applied on the results. For instance, the assistant of a head of department that we consider as the information watcher can present a summary to his head of department.
- **Interpretation:** this means in general, the possibility of the user of the system being able to make the right decisions. It does not mean that the sole user of the system is the decision maker; it can include the information watcher. One can see then the interest in capturing the profile of the decision maker in a metadata stored on the data warehouse which can be used to build a specific data mart for a group of decision makers or even better a particular user.

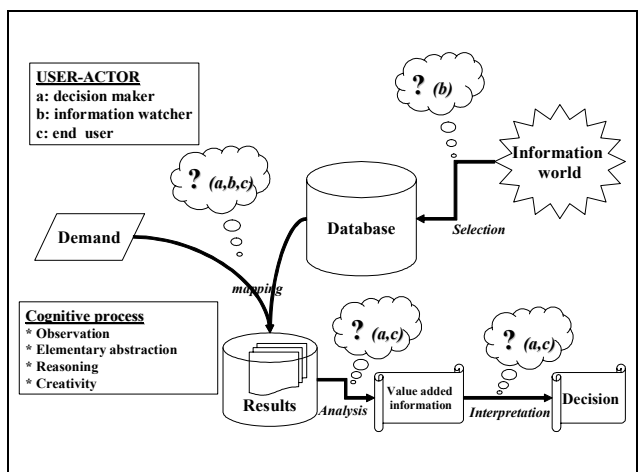

**Figure 2:** A Business Intelligence system.

---

[1] Modélisation et Développement des Systèmes d'Intelligence Economiques – Laboratoire Lorrain de Recherche en Informatiques et ses Applications Nancy, France

Also in this process one can identify three main actors:

- **Decision maker:** this is the individual in the organization that is capable of identifying and posing a problem to be solved in terms of stake, risk or threat that weighs on the organization. In other words, he knows the needs of the organization, the stakes, the eventual risks and the threats the organization can be subjected to.
- **Information watcher:** this refers to the person within the organization that specializes in the methods of collection and analysis of information. His objective is to obtain indicators (using information) or value added information that the decision makers depend on for his decision process. After receiving the problem to be solved as expressed by the decision maker, the information watcher must translate it into information attributes to be collected and which are used to calculate the indicators.
- **End user:** this is the final user of the system; it can be either of the previously outlined users or neither of the two. This user is defined depending on which layer of the business intelligence system he interacts with.

## 2. Modelling User

### 2.1. The role of the user

The user, who in our context is the decision maker, has a central role to play in a business information system. His ability to use the system efficiently is directly linked to the limit of his knowledge of the system. The first thing to do then will be to evaluate his knowledge of the system, use this knowledge to establish the importance of his role, his working habits, the most frequently used data etc.

Next, using this information, a personalised structure can be generated to improve his use of the system. A complete and robust work environment can enormously increase his efficiency. On the other hand, he can bring out the critical elements of the system, the errors, faults and missing points of the system. For this user – decision maker, the decision making process begins by acknowledging a decisional problem, that can be translated as a decisional need. The resolution of a decisional need consists in identifying the needs necessary for such resolution, be it informational, strategic, human etc. we will be concerned, at this time with only the need in information (informational needs). Informational need can be defined as a function of the user model, his environment and his objectives. Our hypothesis is that a data warehouse is the basis of these SIS. A data warehouse is a subject-oriented, integrated, time-variant and non-volatile collection of data in support of management's decision making process [Inmon, 1995]. This data warehouse gives rise to, by filtering in terms of user profiles (or finally a user model), to data marts. These are the smaller dimensions of the data warehouse designated to a department or a function of the organisation. They are periodically alimented and they rely on a multidimensional view of the data. They are not modifiable by the user.

### 2.2. Our Idea

Most of the existing data warehouse systems were designed using the user model based on his profession [Kimball et al., 1998 and Haynes, 2001]. Whereas this model was not complete since each user reacts differently according to his needs and his working habits. For instance, a user/decision maker may have a need that is specific to him (in terms of his personality traits, cognitive style, preferences etc.) [Bouaka and David, 2003] that was not treated in the base. We are trying to respond to a the following question: "what are the parameters that should be added to a user model or which data are to be included in the data warehouse or the data marts that will help the system respond better to his informational needs?".

### 2.3. The proposed user model

The objective of user model is to be able to personalize the responses of the system. User modelling is the way a user and his behaviour are represented. Three categories of model were earlier proposed:

- **User profile:** where to a user is associated the requests that expresses his needs. In this context, the user need is relatively stable. The profile is applied to new information in order to be able to propose to him the information that is more pertinent to him.
- **Implicit user model:** where the behaviour or the preferences of the user is determined in an implicit manner. For instance the visualisation of a document can be interpreted as it being adequate response to his request.
- **Explicit user model:** where the behaviour or the preferences of the user are represented but according to the specifications of the user. Following the earlier example, this user will not only need to open the document but he has to indicate that its degree of pertinence.

Earlier works by [Thiery and David, 2002] on personalization of responses in Information Retrieval Systems (IRS) adapted the four cognitive phases in the human learning process i.e.:

- **Observation phase:** here, the learner gathers information about his environment by observation.
- **Elementary abstraction phase:** the learner describes the objects observed using words, this corresponds to a phase of acquiring the vocabulary of the system being observed.
- **Reasoning and symbolization:** the learner starts to use the vocabulary acquired which implies a higher level of abstraction.
- **Creativity phase:** here the learner discovers and uses the knowledge that were not explicitly presented in the system.

This was transformed into a user model in an IRS context. The first two phases were compressed into **exploration** and this gives a model:
M = {Identity, Objective, {Activity} {Sub-sessions}}

Where;

Activity = {Activity-type, Classification, Evaluation}
Type = {Exploration, Request, Synthesis}
Classification = {Attributes, Constraints}
Evaluation = {System's solution, Degree of Pertinence}

- **Identity:** the identity of the user. This allows the individualisation of the historic of the sessions of the user.
- **Objective:** the principal objective or the real need of the user for the session.
- **Activity:** A user activity that leads to the resolution of his information need. A session is composed of many activities and each activity is defined by three parameters: activity-type, classification, and evaluation.
- **Activity-type:** the types of activity correspond to the different phases of evocative user habits which are in this case exploration, request and synthesis.
- **Classification:** this is the approach we use to access stored information. The classification technique permits the user to express his information requirement in terms of the evocative phases that we are implementing. The user will be able to specify the attributes of the documents to classify and the constraints that are to be met by these documents.
- **Evaluation:** the user will be able to evaluate the pertinence of all the solutions proposed by the system. It should be noted that this evaluation relies on the degree of pertinence and the reasons for this judgement.
- **Sub-sessions:** a sub-session is represented exactly like a main session. The only difference is that the objective of the sub-session is associated to the objective of the main session and a sub-session will not constitute a session apart.

This user model permits the proposition of an information system architecture that relies on a cognitive user evolution. The user can: explore the information base to discover its contents; formulate his requests; add annotations; and link his information retrieval activities to a definite predetermined objective.

### 2.4. Information need

The information need of a user is a concept that varies in definition, according to different researchers and according to the different users [Campbell and Rijsbergen, 1996] and [Xie, 2000]. We believe that the information need of a user is an informational representation of his decisional problem [Goria and Geffroy, 2004]. Defining a decisional problem implies certain level of knowledge on the user and his environment. Therefore, a decisional problem is a function of the user model, his environment and his objective.

We base our definition on that of [Bouaka and David, 2003] where a decisional problem is defined as

$P_{decisional}$ = f(Stake, Individual Characteristics, Environmental Parameters)

Stake is what the organization stands to loose or gain. It is made up of Object, Signal and Hypothesis.

Individual Characteristics refer to the user, his behaviours and his preferences. This includes his Cognitive Style, his Personality Traits, and his Identity.

Environmental parameters mean the input of the society on the organisation. This can be Immediate or Global.

Therefore, a decisional problem can be further broken down into:

$P_{decisional}$ = f((O, S, H), (CS, PT, I), (GE, IE))

Where;

O = the environment **Object** detected by the decision maker;
S = the associated **Signal** i.e. the meaning the decision maker gives to it;
H = the assumption or **Hypothesis** i.e. the possible results or outcome associated with each signal;
CS= **Cognitive Style**, this shows the individual differences in humans;
PT = **Personality Trait**, this is a set cognitive and affective structures conserved with time by individuals to facilitate their adjustment to events, people and decisions;
I = **Identity** is use to reference each individual user and to individualise sessions;
GE = **Global Environment,** this regroups the social, political, economic environment i.e. the image of the organisation;
EI = **Immediate Environment**, this affect the organisation in a direct way, it can include the customers, the suppliers, the competitors etc.

## 3. Application domain

We are limiting ourselves, in the first instance, to an application framework in information retrieval, using a base of documents published by researchers in a research centre. This data warehouse contains publications, historicized and grouped according to the habitual bibliographic nomenclature, of members of the research centre. We had worked on the classification, normalization and improvement of such electronic document resource and our objective is to constitute a real data warehouse of documents from which we could create all type of information analysis. In particular, we propose producing different data marts for the different users of the system.

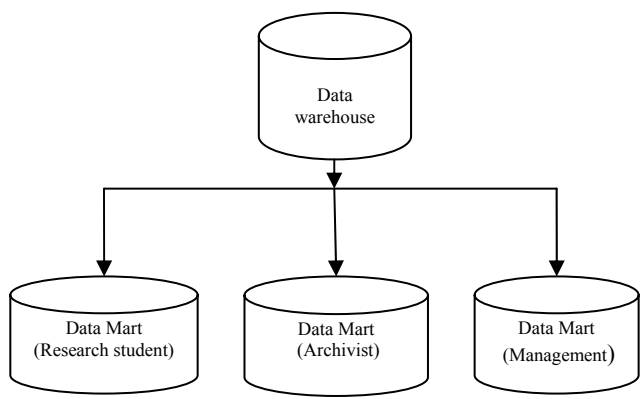

**Figure 3:** Data Marts from the data warehouse of documents.

For instance, we cite examples of data marts obtained during an analysis on this data warehouse of documents. We are no longer searching for the documents themselves but rather we are interested in the information on these documents that can aid in the decision making process. In the figure 3, one can see three data marts derived from the data warehouse on three essential types of users:

- The first contains documents and a representation of the evolution of the researches carried out by users according to research topic (this can be very interesting to a research student who wants to know "who published what?" or the trend of the publications of an author over the years or just the recent ones;

- The second is a base of documents that helps in following the evolution of users' demands for document with time, this can help in identifying a user's research interest in order to be able to propose to him new documents that are related to his research interests. It will also guide against recommending the same document(s) many times over. This will particularly please the archivists or the librarian;

- The third contains information about evolution of the publications according to the different research teams in the laboratory. This will be of importance to the leadership of the centre who can use this as an instrument of evaluation of each research team or even as instrument for deciding the research orientations of the centre.

This supposes that each of these users has a different view of the data from the data warehouse and would want that to him be proposed data that essentially respond to his needs.

It should also be noted that the actual structure of data warehouses does not permit an easy evolution. Thus, during a session of query implementation, we wanted to follow the evolution of publications of each research team. However, we found out that the attribute "EQUIPE[2]" that could have helped us in this regard was missing from the list of attributes. We then ascertain that some attributes necessary for the decision making process could be missing from the base. It is our intention to detect such anomalies and then propose a solution to it.

## 4. Further propositions

In order to resolve the issue of missing attribute(s) or missing data values for a particular attribute we started by classifying the type of information that could be missing.

There are two main types of important attribute that can be missing for the present base:

- User attributes: these are attributes that describe the user, his preferences, his work habits, his needs in information;
- Document attributes: these are the attributes that describe the document contained in the database. They are necessary for the description of the documents.

For each of these attribute types, there are associated data and these associated data are the values that these attributes can take. Therefore it is possible that the attributes are present, but the values are missing. We could also have a situation where both attribute and its associated values are missing.

Depending on which of these parameters is missing the solution to be proposed varies. The premier proposition then is to extend the data sources to not only the existing databases in the documentation centre. For instance, the telephone directory of the laboratory which contains the names of all the staff of the research centre according to their research team can be a source of information to supply the staff names with their respective affiliations. Also the human resources management system can be a good source of information. What we are proposing is the extension of the data sources to all system that could be of help in resolving our needs depending on the missing information.

Secondly, the metadata resident on the data warehouse should also be formulated in such a way as to accommodate the user, his behaviours, and his preferences etc so as to be able to decompose the data warehouse into views (data marts) that are more interesting to each user and also be able to aid the user in his daily use of the data warehouse.

## 5. Conclusion

This model of an information need can be used along with the user model in an IRS context as defined above to further extract a lot of information on the user, his behaviours and why he behaves the way he does when in direct contact with the system.

---
[2] EQUIPE is the French word that translates to TEAM in English. This is used to denote research team in the database (the data actually contained in the base we are using is in French).

Our next phase of the research is condensing these two models into one which will serve as the metadata for reconceptualising of the actual base.

All of the inclusion mentioned above to the information system on document will permit a better personalisation of the system and reduce the time spent by users to get information.

Métamorphoses des organisations Nancy Novembre 2004.

[Xie, 2000] Hong Xie, "Patterns between interactive intentions and information-seeking strategies" in Information Processing and Management 38 (2002). p 55 – 77.